\documentclass[preprint,preprintnumbers]{revtex4}

\usepackage{graphicx}

\usepackage{amssymb}

\usepackage{mathrsfs}

\begin{document}

\noindent

\title{Thermal conductance  of   zero modes  on the surface boundary   of a Weyl semimetal }

\affiliation{Physics Department, City College of the City University of New York,  
New York, New York 10031, USA}
 
\author{D. Schmeltzer}

 \begin{abstract}
Thermoelectric conductance of Dirac materials and in particular zero modes reveals the effect of  topology  .Weyl semimetals with a boundary at $z=0$  give rise  to chiral zero modes  without backscattering  resulting in  a significant contribution  to thermal conductivity.
By  doping  the surface  with paramagnetic impurities  backscattering  is allowed, and the thermal conductivity is controlled by the decrease of the  transmission   function $  | t|^2<1$ .
We attach a thermal reservoir at the edge of the sample   and study the thermal and electrical conductance.  
For  the  ballistic and mesoscopic situations, quantum fluctuations   causes  oscillations of  the thermal and electric conductance. The thermoelectric conductance  varies  periodically with the  voltage bias. 
We compare  the thermal conductance  with and without  impurity scattering  and observe the effects of topology.  An experimental set-up is proposed to test this theory.
 
\end{abstract}

\maketitle

\textbf{I. INTRODUCTION}

Thermal conductance  is the flow of heat that results  from a   temperature gradient\cite{Mahan}. Thermoelectrics are used in cooler  refrigerators based on the Peltier \cite{Goldsmith} effect which predicts the appearance of a heat current when an electric current passes through a material.  Alternatively, the Seebeck effect  generates an electric  current from a temperature gradient \cite{Kamran}. According to \cite{Mahan}, the presence of a disorder  can might enhance the figure of merit \cite{Kamran}. These results have been obtained within  the semiclassical Boltzmann theory. Recent experiment performed by  \cite{Pepper} suggest  that interference effects  are important and may invalidate the   Boltzmann theory.
It was shown that edge states affect the thermal cconductance \cite{Banerjee}.

Another situation where edge modes contribute to the thermal conductance are surface zero modes of    Weyl  semimetals \cite{David}.The Weyl semimetals are topological materials that are protected against localization. The  disordered Weyl semimetals resemble   a  three dimensional Anderson metal \cite{Altland}.  In agreement with this  we find that the boundary surface support zero modes .  As a function of the surface width in the $x$ direction  we find $2N$ chiral modes which propagate in the $y$ direction. Due to  topology the chiral modes are protected against backscattering and the thermal conductance  is given  by $\kappa=2N \frac{\pi^2K^{2}_{B}T}{3h}$  with the value of $ N$ being determinate   by the  temperatur.
  Doping the Weyl surface with paramagnetic impurities  generates a system of $2N$ chiral  backscattering  pairs.
The computation is done by  the applying  Landau Buttiker theory   \cite{Buttiker,Flensberg}  to Dirac materials with particles and anti-particles.
 We obtain  a system  of $N$ one dimensional  zero modes  where the Landau Buttiker theory \cite{Butcher} will be applied. We investigate the thermal effects in the temperature range where the electronic systems obey the ballistic or the  mesoscopic conditions. We find that the thermoelectric conductance fluctuates strongly with the change of the chemical potential . This result is in agreement with the fluctuations  controlled by  density change observed by \cite{Pepper}.

The plan  of this paper is  outlined  as follows:In chapter II we will consider   a Weyl semimetal crystal  with a boundary at $z=0$. As a result, zero modes lie on the two dimensional  boundary at $z=0$.In section III we show that if we restrict the crystal to  a width of  $ D_{x}$ perpendicular to the line which connects the monopole  and the anti- monopole we obtain N pairs of chiral zero modes .  In chapter  IV
we consider the Landauer Buttiker formulation  for   Dirac  fermions  with the perfect transmission  $|t|^2=1$ and    compute the  electric conductance $ G$,thermoelectric conductance $L$ and thermal conductance $\kappa$ at finite temperatures for the ballistic condition.  In chapter V, we include magnetic impurities which   give rise to backscattering    with non perfect transmission,  $|t|^2<1$  allowing to investigate  the mesoscopic conditions  In chapter VI, we propose   an experimental set-up   for testing  th theory. In section VII we present our conclusion.

\vspace{0.2 in}

\textbf{II.  The Weyl  Hamiltonian  with a boundary  at  $z=0$,   confined to the crystal  region $ -L\leq z \leq 0$ }

\vspace{0.2 in}

The bulk of the Weyl semimetals ( $WSMs$)  is dominated by Weyl points and   linear,  low energy  excitation. The Weyl points come in pairs with opposite chirality \cite{Ninomya}.
The surface state of the $WSMs$  is  characterized by $^{''}$Fermi arcs$^{''}$ that link the projection of the bulk Weyl points  in the  Brillouine zone.
The $WSMs$  exist in materials where time-reversal symmetry or inversion are  broken. 
Recently, the non-centrosymmetric and non-crystal magnetic transition-metal monoarsenide/posphides  $TaAs$ ,$TaP$, $NbAs$ and $NbP$ have been predicted to be  $WSMs$ with $12$ pairs of Weyl points \cite{Science}.
We consider  a   $WSM$  Hamiltonian   which describes fermions with opposite chirality  and  two singularities at  $k_{x}=\pm M$.
A $WSM$ model  without a boundary and  two nodes $\vec{M} =[ \pm M ,0,0]$ is given by the Hamiltonian:
\begin{equation}
\tilde{H}=\int\,d^3x \hbar v\Big[\hat{\Psi}^{\dagger}_{R}(\vec{x})\vec{\sigma}\cdot\Big(-i\vec{\partial}-\vec{M}\Big)\hat{\Psi}_{R}(\vec{x})
-\hat{\Psi}^{\dagger}_{L}(\vec{x})\vec{\sigma}\cdot\Big(-i\vec{\partial}-(-\vec{M})\Big)\hat{\Psi}_{L}(\vec{x})\Big]
\label{equattion}
\end{equation}
This model is oversimplified and does not include the  band dispersion which     connects  the  two nodes. In order to observe this connection,  we need  to study  a model  with two  non-linearly dispersed   bands. We are guided by the fact that  the singularities at  $k_{x}=\pm M$ describe a monopole and   anti-monopole . 
 To describe the crossing of the bands in momentum space, we will introduce a quadratic function of momentum   $g(k^2_{x}-M^2)$  which  reproduces  the nodes at $ \pm M$ ( this polynomial is obtained by replacing $-cos(k_{x}) +1 \approx \frac{k^2_{x} }{2}$) for   the two  band  Hamiltonian   $\hat{h}(\vec{k},z)$:
$\hat{h}(\vec{k},z)=\hbar v \Big[\sigma_{y}\tau_{3}k_{y}+\sigma_{z}\tau_{3}k_{z}+\sigma_{y}\tau_{2} g(k^2_{x}-M^2)\Big]$.
\noindent
The Hamiltonian with  the boundary at $z=0$  and  potential energy  $\hbar vk_{0}$ is given by:
\begin{eqnarray}
&&H=\int\frac{d^{2}k}{(2\pi)^2} \int_{-L}^{0}\,dz\Big[\hbar v\hat{\Psi}^{\dagger}(\vec{k},z)h(\vec{k},z)\hat{\Psi}(\vec{k},z)\Big];
h(\vec{k},z)=h^{(0)}(\vec{k},z)+h^{(1)};(\vec{k})\nonumber\\&&
h^{(0)}(\vec{k},z)= \Big(\sigma_{z}\tau_{3}(-i\partial_{z})+\sigma_{y}\tau_{2} g(k^2_{x}-M^2)\Big)\hspace{0.1 in};
h^{(1)}(\vec{k})= \Big(\sigma_{y}\tau_{3}k_{y}-k_{0}I\Big)\nonumber\\&&
\end{eqnarray}
 with the zero modes solution, $h^{(0)}(\vec{k},z)U_{i=1,2}(\vec{k},z)=EU_{i=1,2}(\vec{k},z)$.
\begin{equation}
 U(\vec{k},z)  =e^{\lambda z}V(\vec{k})=\sum_{s=\pm}\Big[\theta[k^2_{x}-M^2]e^{g(k^2_{x}-M^2)z}\eta_{i,+;s}+\theta[-k^2_{x}+M^2]e^{-
g(-k^2_{x}+M^2)z}\eta_{i,-;s}(\vec{k})\Big]; \vspace{0.05 in} k^2\neq M^2 
\label{eqw}
\end{equation}
We mention that in addition to the zero modes $E=0$ we have non-zero modes excitations. At  low temperatures, we can  neglect the nonzero modes. Therefore for temperatures $T<T_{M}$ we ignore the non zero modes.
The probability for exciting non zero  modes  on the surface is  $ f_{f.D.n\neq 0}=(1+e^{\frac{\hbar v \sqrt{k^2_{y}+k^2_{x}}-\hbar v k_{0}}{K_{B}T}})^{-1}$ and the probablity for  exciting zero modes is $ f_{f.D.n= 0}=(1+e^{\frac{\hbar v |k_{y}|-\hbar v k_{0}}{K_{B}T}})^{-1}$. At low temperatures we have , $ f_{f.D.n\neq 0}<< f_{f.D. n=0}$.

According to \cite{Witten}, we  identify  the zero modes with boundary surface  states at $ z=0$.
The   zero mode solutions $\eta_{i,\pm;s}$, $i=1,2$ with the index $\pm$ refering  to the momentum  region $k^2_{x}>M^2$ or $M^2>k^2_{x}$ while $s$  refers  to $\pm k_{x}$ \cite{David}.
\begin{eqnarray}
&&\eta_{1,+;s}=\sqrt{\frac{1}{2}}\Big[i ,0,0,1\Big]^{T} ,\hspace{0.05 in} \eta_{1,-;s}=\sqrt{\frac{1}{2}}\Big[-i,0,0,1\Big]^{T} \nonumber\\&&
\eta_{2,+;s}=\sqrt{\frac{1}{2}}\Big[0,i,1,0\Big]^{T}, \hspace{0.05 in}   
\eta_{2,-;s}=\sqrt{\frac{1}{2}}\Big[0,-i,1,0\Big]^{T}\nonumber\\&&
\end{eqnarray}

\noindent
Following \cite{David,} we  diagonalyze  $h^{(1)}(\vec{k})$ in terms of the zero modes operators  $C_{1,-;s}$, $C_{2,-;s}$,   $C_{1,+;s}$, $C_{2,+;s}$ .
\begin{eqnarray}
&&H^{\perp}=\int\frac{dk_{y}}{(2\pi)}\sum_{k_{x}}\sum_{\pm}  \Big[\hbar v \Big[ k_{y}\Big(-iC^{\dagger}_{1,-;s}(\vec{k})C_{2-;s}(\vec{k}) +iC^{\dagger}_{2-;s,}(\vec{k})C_{1-;,s}(\vec{k})\Big) \nonumber\\&&-k_{0}\Big(C^{\dagger}_{1,-;s}(\vec{k})C_{1,-;,s}(\vec{k})+C^{\dagger}_{2,-;s}(\vec{k})C_{2,-;s}(\vec{k})\Big)\Big]\theta[k^2_{x}-M^2]   \nonumber\\&& + \hbar v \Big[ k_{y}\Big(-iC^{\dagger}_{1,+;s}(\vec{k})C_{2,+;s}(\vec{k}) +iC^{\dagger}_{2,+;s,}(\vec{k})C_{1,+;,s}(\vec{k})\Big) \nonumber\\&&-k_{0}\Big(C^{\dagger}_{1,+;s}(\vec{k})C_{1,+;,s}(\vec{k})+C^{\dagger}_{2,+;s}(\vec{k})C_{2,+;s}(\vec{k})\Big)\Big]\theta[-k^2_{x}+M^2]\Big]\nonumber\\&&
\end{eqnarray}
  We replace  the  operator $C_{1,-;s}$, $C_{2,-;s}$,   $C_{1,+;s}$, $C_{2,+;s}$ with the   eigenvalue  operators,  $C_{\pm,\pm;s}$.  We have the transformation : $C_{1,-;s}= \frac{1}{\sqrt{2} }(C_{+,-;s}+C_{-,-;s})$ ,  $C_{1,+;s}= \frac{1}{\sqrt{2} }(C_{+,+;s}+C_{-,+;s})$  $C_{2,-;s}=\frac{i}{\sqrt{2} }(C_{+,-;s}-C_{-,-;s})$ ,$C_{2,+;s}= \frac{1}{\sqrt{2} }(C_{+,+;s}+C_{-,+;s})$.   We replace the spinors $ \eta_{i,+;s}$ ,$ \eta_{i,-;s}$ $i=1,2$ with  transformed  spinors $\eta_{\pm,+;s}$, $\eta_{\pm,-;s}$:
\begin{equation}
\eta_{+,-;s}=\frac{1}{\sqrt{2} }[-i,1,i,1],\hspace{0.01in}\eta_{-,-;s}=\frac{1}{\sqrt{2} }[-i,-1,-i,1];
\eta_{+,+;s}=\frac{1}{\sqrt{2} }[i,-1,i,1],\hspace{0.01in}\eta_{-,+;s}=\frac{1}{\sqrt{2} }[i,1,-i,1]
\label{spinors}
\end{equation}
For a sample of width $D_{x}$, the momentum $k_{x}$ is discreet, $k_{x}=\frac{2\pi}{D_{x}}(n_{x}+\frac{1}{2})$, $n_{x}=\pm1,\pm2....$   (see Figure.1) with  $ n_{x,max.}$  given by  $ n_{x,max.}=\frac{M}{\frac{2\pi}{D_{x}}}-\frac{1}{2}$ . At finite temperature  $T<T_{M}$, we can define $N$ as  the number of  excited modes  in the $ x$ direction, $\mathbf{ N=\frac{K_{B}T}{\hbar v_{\perp}  \frac{2\pi}{D_{x}}}}$  ($ \hbar v_{\perp} \frac{2\pi}{D_{x}}$ the excitation energy in  the $x$ direction which satisfies the  conditions $ v_{\perp}<v$) .When $n_{x,max.} $ is larger than the thermal mode excited $N$, we have the representation ( $n_{x,max.}>N $):
\begin{eqnarray}
&&H^{\perp}=\sum_{s=\pm}\int\,\frac{dk_{y}} {2\pi}\Big[\hbar v (k_{y}-k_{0})\Big(C^{\dagger}_{+,-;s}(\vec{k})C_{+,-;s}(\vec{k})+
C^{\dagger}_{-,-;s}(\vec{k})C_{-,-;s}(\vec{k})\Big)\nonumber\\&&+\sum_{n_{x}=1}^{N}\sum_{s=\pm}\int\,\frac{dk_{y}}{2\pi}\hbar v (k_{y}-k_{0})\Big(C^{\dagger}_{+,-;s,n_{x}}(\vec{k})C_{+,-;s,n_{x}}(\vec{k})+
C^{\dagger}_{-,-;s,n_{x}}(\vec{k})C_{-,-;s,n_{x}}(\vec{k})\Big)\nonumber\\&&+\sum_{n_{x} =N}^{n_{x,max}}\sum_{s=\pm}\int\,\frac{dk_{y}}{2\pi}\hbar v (k_{y}-k_{0})\Big(C^{\dagger}_{+,-;s,n_{x}}(\vec{k})C_{+,-;s,n_{x}}(\vec{k})+
C^{\dagger}_{-,-;s,n_{x}}(\vec{k})C_{-,-;s,n_{x}}(\vec{k})\Big)\Big]\nonumber\\&&
\end{eqnarray}
For the  case  $n_{x,max.}>N $ we have only    modes which obey  $ k^2_{x}>M^2$
 with  spinors $\eta_{+,-;s}$ and $\eta_{-,-;s}$.

In the opposite situation,  $n_{x,max.}<  N$, we have two type of modes :  modes with $ k^2_{x}>M^2$  and modes with  $ k^2_{x}<M^2$.  For this case, we have both  spinors $\eta_{+,+;s}$ , $\eta_{-,+;s}$ see Eq.$3$ and $\eta_{+,-;s}$ , $\eta_{-,-;s}$.
\begin{eqnarray}
&&H^{\perp}= \sum_{n_{x}=1}^ {n_{x,max.}}\sum_{s=\pm}\int\,\frac{dk_{y}}{2\pi}\hbar v (k_{y}-k_{0})\Big(C^{\dagger}_{+,-;s,n_{x}}(\vec{k})C_{+,-;s,n_{x}}(\vec{k})+
C^{\dagger}_{-,-;s,n_{x}}(\vec{k})C_{-,-;s,n_{x}}(\vec{k})\Big)\nonumber\\&&+\sum_{n_{x}=n_{x,max.}}^{N}\sum_{s=\pm}\int\,\frac{dk_{y}}{2\pi}\hbar v (k_{y}-k_{0})\Big(C^{\dagger}_{+,+;s,n_{x}}(\vec{k})C_{+,+;s,n_{x}}(\vec{k})+C^{\dagger}_{-,+;s,n_{x}}(\vec{k})C_{-,+;s,n_{x}}(\vec{k})\Big)\nonumber\\&&
\end{eqnarray}
 For this case we use the spinors $\eta_{+,+;s}$ and $\eta_{-,+;s}$ (see Eq.$3$).

In Figure $1$, we show the one dimensional channels  as a function of the discreete momentum $k_{x}$. We observe that for each value of  $k_{x}$ , we have left and right  fermions. This demonstrates that the surface of Weyl semimetal is equivalent to $N$ pair of chiral fermions  (see Figure.1) .

\vspace{0.2 in}

\textbf{III-The quasi one-dimension edge mode}

\vspace{0.2in}

For a narrow width, $D_{x}$, we can consider  only  the zero mode  at fixed momentum  $ k_{x}=\pm\frac{\pi}{D_{x}}$ which propagates  along the $ y$ direction. We have   a mode with two chiralities. 
The $y$  direction needs to be  perpendicular to the line connecting the  pair of  monopole anti-monopole. Thus, we replace  $H^{\perp}$ $\rightarrow$ $H^{edge}$.
The Hamiltonian at the fixed momentum $ k_{x}=\pm\frac{\pi}{D_{x}}$ takes the form:
\begin{equation}
H^{edge}=\sum_{s=\pm}\int\,\frac{dk_{y}}{2\pi}\Big[\hbar v (k_{y}-k_{0})C^{\dagger}_{+,-;s}(\vec{k})C_{+,-;s}(\vec{k})
+ \hbar v (-k_{y}-k_{0}) C^{\dagger}_{-,-;s}(\vec{k})C_{-,-;s}(\vec{k}) \Big]
\label{1D}
\end{equation}
We represent $ C_{+,-;s}(k_{y})$ and  $C_{-,-;s}(k_{y})$ in terms of the particle operators  $a(k)$ and anti-particle $b^{\dagger}(k)$:The second minus stand for the case  that $ k^2_{x}<M^2$ .
\begin{equation}
C_{+,-;s}(k_{y})=a_{+,-;s}(k_{y})\theta[k_{y}]+b^{\dagger}_{+,-;s}(-k_{y})\theta[-k_{y}];\hspace{0.05 in} C_{-,-;s}(k_{y})=a_{-,-;s}(-k_{y})\theta[-k_{y}]+b^{\dagger}_{-,-;s}(k_{y})\theta[k_{y}]
\label{coefficients}
\end{equation}
For $0\leq k<\infty$, we introduce $k=q+k_{F}$,  and for  $-\infty<k\leq0$, we introduce $k= q -k_{F}$ with
 $- 2k_{F}\leq q\leq 2 k_{F}$. We introduce an ultraviolet cut-off $\Lambda$ , $2k_{F}=\frac{\pi}{a}\approx \Lambda$.
At half filling  $k_{0}=0$, the   fermion field is given in terms of the left and right movers : $C_{-,-;s}(y)=e^{-ik_{F}y}\psi_{-,-;s}(y)\eta_{-,-;s}$  , $C_{+,-;s}(y)=e^{ik_{F}y}\psi_{+,-;s}(y)\eta_{+,-;s}$ , where :
\begin{eqnarray} 
&&\psi_{+,-;s}(y)=\int_{-\Lambda}^{\Lambda}\frac{d q}{2\pi}\psi_{+,-;s}(q)\eta_{+,-;s}e^{iqy}\hspace{0.01 in}; \psi_{-,-;s}(y)=\int_{-\Lambda}^{\Lambda}\frac{d q}{2\pi}\psi_{-,-;s}(q)\eta_{-,-;s}e^{-iqy}\nonumber\\&&
\psi_{+,-;s}(q)=\alpha_{+,-;s}(q)\theta[q] +\beta^{\dagger}_{+,-;s}(-q)\theta[-q]\hspace{0.01in};\psi_{--;s}(q)=\alpha_{-,-;s}(q)\theta[q] +\beta^{\dagger}_{-,-;s}(-q)\theta[-q]\nonumber\\&&
 \alpha_{+,-;s}(q)=a_{+,-;s}(q+k_{F}), \beta^{\dagger}_{+,-;s}(-q)=b^{\dagger}_{+,-;s}(-q-k_{F});\hspace{0.01 in}\alpha_{-,-;s}(q)=a_{-,-;s}(q+k_{F})\nonumber\\&& \beta^{\dagger}_{+,-;s}(-q)=b^{\dagger}_{+,-;s}(-q-k_{F})\nonumber\\&&
\end{eqnarray}
 For $ k_{0}\neq0$  we find following the representation  for the Hamiltonian in Eq.$(4)$:
\begin{eqnarray}
&&H^{edge}_{+}=\sum_{s=pm}\int_{-\infty}^{\infty}dy h^{edge}_{+;s}(y)=
\nonumber\\&&\sum_{s=\pm}\int_{-\infty}^{\infty}\frac{dk_{y}}{2\pi}C^{\dagger}_{+,-;s}(k_{y})C_{+,-;s}(k_{y})\hbar v(k_{y}-k_{0})\approx
\int_{0}^{\Lambda}\frac{dq}{2\pi}\Big[\hbar v(q-V)\alpha^{\dagger}_{+,-;s}(q)\alpha_{+,-;s}(q)\nonumber\\&&-\hbar v(q+V)\beta_{+,-;s}(q)\beta^{\dagger}_{+,-;s}(q)\Big]=\sum_{s=\pm}\int_{-\infty}^{\infty}dy\psi^{\dagger}_{+,-;s}(y)(\partial_{y}-V)\psi_{+.-;s}(y) , \hspace{ 0.01in} V=k_{0}-k_{F}\nonumber\\&&
H^{edge}_{-}=\sum_{s=\pm}\int_{-\infty}^{\infty}dy h^{edge}_{-;s}(y)=
\nonumber\\&&\int_{-\infty}^{\infty}\frac{dk_{y}}{2\pi}C^{\dagger}_{-,-;s}(k_{y})C_{-,-;s}(k_{y})\hbar v(-k_{y}-k_{0})\approx\sum_{s=\pm}
\int_{0}^{\Lambda}\frac{dq}{2\pi}\Big[\hbar v(q-V)\alpha^{\dagger}_{-,-;s
}(q)\alpha_{-,-;s}(q)\nonumber\\&&-\hbar v(q+V)\beta_{-,-;s}(q)\beta^{\dagger}_{-,-;s}(q)\Big]\ =\int_{-\infty}^{\infty}dy\psi^{\dagger}_{-,-;s}(y)(-i\partial_{y}-V)\psi_{-,-;s}(y), \hspace{ 0.01in} V=k_{0}-k_{F}\nonumber\\&&
\end{eqnarray}
We notice the negative sign for the anti-particle Hamiltonian  and the effect of the voltage bias  $V=k_{0}-k_{F}$   which represents the  chemical potential shift .

Using the continuity equation  for the left and right movers  $\partial_{t}(e\psi^{\dagger}_{-,-:s}(y)\psi_{-,-;s}(y))+\partial_{y}J^{el}_{-;s}(y)=0$, $\partial_{t}(e\psi^{\dagger}_{+;s}(y)\psi_{+;s}(y))+\partial_{y}J^{el}_{+;s}(y)=0$, we obtain the electrical currents : $ J^{el}_{-;s}(y)=ev\psi^{\dagger}_{-,-;s}(y)\psi_{-,-;s}(y)$ and $ J^{el}_{+;s}(y)=-ev\psi^{\dagger}_{+,-;s}(y)\psi_{+,-;s}(y)$ 

 Due to the fact that the energy is computed relative  to the Fermi energy allows us to identify the thermal energy with the energy computed from the Hamiltonian  with the subtracted ground state. As a result the heat current and the energy current  are   the same.
The heat current is obtained from the continuity equation for the energy densities  $h^{edge}_{+;s}(y)$ and  $h^{edge}_{-;s}(y) $:
\noindent
\begin{eqnarray}
&&\frac{d h^{edge}_{-;s}}{dt}+\partial_{y}J^{heat}_{-;s}(y,t)=0; \frac{d h^{edge}_{+;s}}{dt}+\partial_{y}J^{heat}_{+;s}(y,t)=0 \nonumber\\&& 
 J^{heat}_{+;s}(y,t)=-\hbar v^2\psi^{\dagger}_{+,-;s}(y)(i\partial_{y}-V)\psi_{+,-;s}(y)\hspace {0.01in}; J^{heat}_{-;s}(y,t)=\hbar v^2\psi^{\dagger}_{-,-;s}(y)(-i\partial_{y}-V)\psi_{-,-;s}(y)\nonumber\\&&
\end{eqnarray}
The thermal current is given by  the expectation value of the operator  $ J^{heat}_{\pm;s}(Q\rightarrow 0,\omega\rightarrow 0 )$
\begin{eqnarray} 
&& \langle J^{heat;s}_{+;s}(Q\rightarrow 0,\omega\rightarrow 0 )\rangle =-\hbar v^{2}\int_{0}^{\Lambda}\frac{dq}{2\pi}\Big[(q-V) \langle\alpha^{\dagger}_{+,-;s}(q))\alpha_{+,-;s}(q)\rangle-(q+V) \langle\beta_{+,-,-;s}(q)\beta^{\dagger}_{+,-;s}(q)\rangle\Big]\nonumber\\&&
\langle J^{heat}_{-;s}(Q\rightarrow 0,\omega\rightarrow 0 )\rangle =\hbar v^{2}\int_{0}^{\Lambda}\frac{dq}{2\pi}\Big[(q-V) \langle\alpha^{\dagger}_{-,-;s}(q))\alpha_{-,-;s}(q)\rangle-(q+V) \langle\beta_{-,-;s}(q)\beta^{\dagger}_{-,-;s}(q)\rangle\Big]\nonumber\\&&
\end{eqnarray}
Where $ \langle\alpha^{\dagger}_{\pm;s}(q))\alpha_{\pm,-;s}(q)\rangle=n_{F}(q-V)=\frac{1}{1+e^{\frac{\hbar vq-V}{K_{B}T}}}$, $\langle\beta_{\pm,-;s}(q)\beta^{\dagger}_{\pm,-;s}(q)\rangle=1-n_{F}(q+V)=1-\frac{1}{1+e^{\frac{\hbar vq+V}{K_{B}T}}}$.

\vspace{0.2in}

\textbf{IV-The electric and  thermal conductivity: an application of the Landau Butikker approach  to one- dimensional Dirac fermions}

\vspace{0.2in}

We will follow the Landau Buttiker approach based on the $\mathbf{S}$ matrix   given by  \cite{Buttiker,Flensberg,Weinberg}  using the modification introduced for particle and anti-particle formulation .

\vspace{0.2in}

\textbf{IVa-The perfect transmission at finite temperatures.}

\vspace{0.2in}

 In the presence of a random potential $U(y)$, the spinor structure of the zero modes Eq.$(6)$ shows that    backscattering     is prohibited. 
As a result, the transmission function is  $|t(\epsilon)|^2=1$.

We will  first consider  the electrical conductance. We attach the left reservoir to a  source with  a voltage $eV_{G}$ and the right reservoir to a source with zero voltage  $V_{G}=0$.
 When we   add  the contribution  from the two leads, we obtain \cite{Flensberg} :
\begin{eqnarray}
&&\sum_{s=\pm} \langle J^{el}_{+;s}(Q\rightarrow 0,\omega\rightarrow 0 )\rangle+ \langle J^{el}_{-;s}(Q\rightarrow 0,\omega\rightarrow 0 )\rangle\nonumber\\&&=\sum_{s=\pm}\int_{0}^{\Lambda}\frac{dq}{2\pi}ev\Big[\Big(n_{F}(q-V+eV_{G})-(n_{F}(q-v)\Big)-\Big(n_{F}(q+V -eV_{G})-n_{F}(q+V)\Big)\Big] \nonumber\\&&\approx
\frac{2e^2}{h}V_{G}\Big[\int_{\frac{-V}{K_{B}T}}^{\infty}dx\partial_{x}n_{F}(x)+\int_{\frac{V}{K_{B}T}}^{\infty}dx\partial_{x}n_{F}(x) \Big], \hspace {0.01 in} x=\frac{\hbar v q}{K_{B}T} \nonumber\\&&
\end{eqnarray}
 The electrical conductance   $G=\frac{2e^2}{h}$ is  depicted in Fig. 3.

Next, we consider the  thermoelectric effects . We attach  a  thermal reservoir  at  temperatures $T+\Delta T$ to the left   and the   reservoir at   temperature  $T$ to the right  \cite{Mahan,Butcher,Bennakker}.
Considering the particle and  anti-particle contributions,  we find the following  equation for the thermal conductivity:
\begin{eqnarray}
&&\sum_{s=\pm} \langle  J^{heat}_{+;s}(Q\rightarrow 0 ,\Omega\rightarrow 0)\rangle+\langle J^{heat}_{-;s}(Q\rightarrow 0,\Omega\rightarrow 0)\rangle=\hbar v^{2}\cdot\nonumber\\&&
\int_{0}^{\Lambda}\frac{dq}{2\pi}\Big[(q-V)\Big(n_{F}(q-V)_{T+\Delta T}-n_{F}(q-V)_{T}\Big)+(q+V)\Big((n_{F}(q+V)_{T+\Delta T}-n_{F}(q+V)_{T}\Big)\Big]\nonumber\\&&\approx 2 \frac{ K^{2}_{B} T}{h}(-\Delta T)\Big[\int_{\frac{-V}{K_{B}T}}^{\infty}  x^2 \partial_{x}n_{F}(x) dx+ \int_{\frac{V}{K_{B}T}}^{\infty} x^2 \partial_{x}n_{F}(x) dx\Big]\nonumber\\&&
\end{eqnarray}
The finding is depicted in Figure $2$, $ \kappa=\frac{2\pi^2K^2_{B}T}{3h}$.

The electric current generated by a thermal gradient is given by:
\begin{eqnarray} 
&& \sum_{s=\pm} \langle J^{el}_{+;s}(Q\rightarrow 0 ,\Omega\rightarrow 0)\rangle+\langle J^{el}_{-}(Q\rightarrow 0,\Omega\rightarrow 0)\rangle\nonumber\\&&=
\int_{0}^{\Lambda}\frac{dq}{2\pi}ev| t|^{2}\Big[\Big(n_{F}(q-V)_{T+\Delta T}-n_{F}(q-V)_{T}- \Big (n_{F}(q+V)_{T+\Delta T}-n_{F}(q+V)_{T}\Big)\Big]\nonumber\\&&
2\frac{eK_{B}}{h}(-\Delta T) \Big[ \int_{-\frac{V}{K_{B}T}}^{\infty} x\partial_{x} n_{F}(x) dx +  \int_{\frac{V}{K_{B}T}}^{\infty} x\partial_{x} n_{F}(x) dx\Big]\nonumber\\&&
\end{eqnarray}
We observe that  the thermoelectric  conductance $ L$ disappears, as  shown in Figure $4$.

The temperature determins $N$,  the number of excited transversal modes. As result  the conductance $G$, the  thermal conductance $kappa$, the thermoelectric conductance  $L$  are replaced by:$G=N\frac{2e^2}{h}$, $\kappa=N\frac{ 2\pi^2}{3h}K^2_{B}T$ and $L=N\frac{2e}{h}K_{B}\cdot 0$.

\vspace{0.2in}

\textbf{IVb- The ballistic conductances for $| t|^2 =1$ }

\vspace{0.2in}

The thermal current for $|t(\epsilon)|^2=1$ for  a finite size system of the size $ L<1\mu $, and  a temperature of $3K$. The energy in the lowest mode is $\epsilon_{0}=0.63 \cdot 10^{-3}ev$ and the energies in the wire are $\epsilon_{n}=(n+\frac{1}{2})\epsilon_{0}$ . For a  bandwidth of $ 0.1 ev$  we will have $100$ modes.  For a finite temperatures we  have $N$ pair of channels.

The  thermal conductance is given by :
\begin{eqnarray}
&& I _{heat}= N\cdot\frac{ v}{L}\sum_{n=1}^{n=160}\nonumber\\&&
\Big[(\epsilon_{n}-V)\Big(n_{F}(\epsilon_{n} -V)_{T+\Delta T}-n_{F}(\epsilon_{n}-V)_{T}\Big)+ (\epsilon_{n}+V)\Big( n_{F}(\epsilon_{n}+V)_{T+\Delta T}-n_{F}(\epsilon_{n}+V)_{T}\Big)\Big]\nonumber\\&&\nonumber\\&&= N\cdot\frac{K_{B}T v}{L}\sum_{n=1}^{n=200}\nonumber\\&&
\Big[(\frac{\epsilon_{n}-V}{K_{B}T})\Big(n_{F}(\epsilon_{n} -V)_{T+\Delta T}-n_{F}(\epsilon_{n}-V)_{T}\Big)+ (\frac{\epsilon_{n}+V}{K_{B}T})\Big( n_{F}(\epsilon_{n}+V)_{T+\Delta T}-n_{F}(\epsilon_{n}+V)_{T}\Big)\Big]\nonumber\\&&
\end{eqnarray}
The heat current  at $T=3K$    in Figure $5$
  fluctuates as a function of the bias voltage $V$ measured in electron volts,
 $I_{heat}=NK_{B} T\frac{v}{L}[0.15-0.25]\frac{ Joule}{sec}$.

The electrical conductance is shown in  Figure $6$. The conductance oscillates as a function of the  voltage bias $V$.
\begin{eqnarray}
&&I_{el.}=N\cdot \frac{ev}{L}
\cdot\sum_{n=1}^{n=160}
\Big[\Big(n_{F}(\epsilon_{n} -V+eV_{G})_{T}-n_{F}(\epsilon_{n}-V)_{T}\Big)-\Big( n_{F}(\epsilon_{n}+V-eV_{G})_{T}-n_{F}(\epsilon_{n}+V)_{T}\Big)\Big]\nonumber\\&&
\end{eqnarray}
The electrical current  is $I_{el}=N\frac{ev}{L}[0.05-0.09]Ampere\approx N \cdot 10^{-7}[0.05 -0.09] Ampere$.

The thermoelectric current is given by:
\begin{eqnarray}
&&I_{el.}=N \cdot \frac{ e v}{L}\cdot \sum_{n=1}^{n=160}
\Big[\Big(n_{F}(\epsilon_{n} -V)_{T+\Delta T}-n_{F}(\epsilon_{n}-V)_{T}\Big)-\Big( n_{F}(\epsilon_{n}+V)_{T+\Delta T}-n_{F}(\epsilon_{n}+V)_{T}\Big)\Big]\nonumber\\&&
\end{eqnarray}
The thermoelectric current in  Figure $7$ shows strong oscillations as  a function of the  voltage  bias $V$,$ I_{el}=N\frac{ev}{L}[\pm0.01]Ampere\approx N \cdot 10^{-7} [\pm 0.01] Ampere$
The results in  Fiigure $7$  are in agreement with the experimental results \cite{Pepper} in silicon  where oscillations for  the Seebeck coefficient   have been reported as a function of the electronic density.

\vspace{0.2in}

\textbf{V-The conductances  for the  mesoscopic conditions}

\vspace{0.2in}

 In order to have a finite scattering matrix element, we need to have backscattering.  For this purpose, we use a model of   a random magnetic field   $(\sigma_{1}\otimes I) H_{x}(y) $ with the property : $\langle H_{x}(q) H_{x}(q')\rangle =H_{sc}\delta[q+q']$.  Such a model is obtained  from  a  one dimensional configuration of paramagnetic impurities.  Doping the Weyl semimetal  surface with  paramagnetic  impurities  in   a narrow strip  of width  $d_{x}<D_{x}$ in the $x$ direction guarantees that the  momentum  scattering will be in the $y$ direction. Due to the  spinor representation in Eq.(4) we find that the  matrix elements  $(\sigma_{x}\otimes I)$ are  non-zero.
  For such a scattering potential we obtain:
\begin{eqnarray}
&& \int_{-\infty}^{\infty}dy\psi^{\dagger}_{+,-;s}(y)(\eta_{+,-;s})^{\dagger}(\sigma_{x}\otimes I)\eta_{-,-;s} H_{x}(y))\psi_{-,-;s}(y)\neq 0\nonumber\\&&\int_{-\infty}^{\infty}dy\psi^{\dagger}_{-,-;s}(y)(\eta_{-,-;s})^{\dagger}(\sigma_{x}\otimes I))\eta_{+,-;s}H_{x}(y)\psi_{+,-;s}(y)\neq0\nonumber\\&&
\int_{-\infty}^{\infty}dy\psi^{\dagger}_{+,+;s}(y)(\eta_{+,+;s})^{\dagger}(\sigma_{x}\otimes I)\eta_{-,+;s} H_{x}(y))\psi_{-,+;s}(y)\neq 0\nonumber\\&&\int_{-\infty}^{\infty}dy\psi^{\dagger}_{-,+;s}(y)(\eta_{-,+;s})^{\dagger}(\sigma_{x}\otimes I))\eta_{+,+;s}H_{x}(y)\psi_{+,+;s}(y)\neq0\nonumber\\&&
\end{eqnarray}
 Due to the spinor structure, $\eta_{\pm,+;s}$ and $\eta_{\pm,-;s}$  the scattering matrix elemets $(\sigma_{x}\otimes I)$ will occour  between the pair with the same  momentum  $k_{x}$. 
 The $\mathbf{S}$  matrix  is identical for holes and particles and  determin the  transmission function $|t(\epsilon)|^2<1$. The transmission function is obtained from the $T$ matrix  formula given in Eq. $5.40$  page $88$ \cite{Flensberg}.
For many channels, for the case   $n_{x max.}>N$  we have : 
\begin{eqnarray}
&&\Psi(x,y)=\sum_{s=\pm}\Big[\Big(e^{ik_{F}y}\psi_{+,-;s,}(y)\eta_{+,-;s} +e^{-ik_{F}y}\psi_{-,-;s,}(y)\Big )\eta_{-,-;s}+\nonumber\\&&\sum_{n_{x}=1}^{N}\Big(e^{ik_{F}y}\psi_{+,-;s,n_{x}}(x,y) )\eta_{+,-;s}+e^{-ik_{F}y}\psi_{-,-;s,n_{x}}(x,y))\eta_{-,-;s}\Big )]\nonumber\\&&
\end {eqnarray}

The  scattering matrix with for a  strip of paramagnetic impurities is given by the field $H_{x}(y)$.This field will generate  backscattering matrix elements.   
Since the scattering potential is only a function of $y$,  the backscattering   will occur  at the same momentum $k_{x}$.

In the case of  $n_{x,max}<N$, we have the representation:
\begin{eqnarray}
&&\Psi(x,y)= \sum_{s=\pm}\Big[\Big(e^{ik_{F}y}\psi_{+,-;s,}(y)\eta_{+,-;s} +e^{-ik_{F}y}\psi_{-,-;s,}(y)\Big )\eta_{-,-;s}+\nonumber\\&&\sum_{n_{x}=n_{x,max}}^{\infty}\Big(e^{ik_{F}y}\psi_{+,-;s,n_{x}}(x,y) \eta_{+,-;s}+e^{-ik_{F}y}\psi_{-,-;s,n_{x}}(x,y))\eta_{-,-;s}\Big )\nonumber\\&&+\sum_{s=\pm}\sum_{n_{x}=n_{x,max}}^{N}\Big(e^{ik_{F}y}\psi_{+,+;s,n_{x}}(x,y) \eta_{+,+;s}+e^{-ik_{F}y}\psi_{-,+;s,n_{x}}(x,y))\eta_{-,+;s}\Big )
\end {eqnarray}
In order to apply the Landau Buttiker theory we need  to be under conditions  where the elastic scattering length $l_{e}$ , the length of the system $L$  and the thermal length $L_{T}$  obey the relations $l_{e}\leq L\leq L_{T}$. 
$l_{e}$ is determined by the disorder.
The use of a random   field in the $x$ direction (or to have magnetic impurities) will generate backscattering.  As a  result for low temperatures we will be able to satisfy the condition  $l_{e}\leq L\leq L_{T}$. 
We will consider a finite size system of the size $ L<1\mu m $, and  a temperature of $3K$.
 Due to  the  $(\sigma_{1}\otimes I) H_{x}(y) $ scattering field $\langle H_{x}(q) H_{x}(q')\rangle =H_{sc}\delta[q+q']$
 we will have localized states in one dimension .   According to  \cite{Weinberg} the $\mathbf{S}$  matrix and the transmission matrix are given in Eqs.$ 3.2.7$ $3.2.8$. Eq.$3.5.3$  \cite{Weinberg} contains an
explicit form of the transmssion function from which we can obtain the transmission function  in the Lorentz approximation at finite temperature: 
\begin{eqnarray}
&&|t(\epsilon)|^2\approx \sum _{n}\frac{(\frac{\Gamma}{2})^2}{(\epsilon -\epsilon_{n})^2+(\frac{\Gamma}{2})^2}\nonumber\\&&
\end{eqnarray}
Using this approximation  we compute the thermoelectric current.
\begin{eqnarray}
&&I_{el.}= N\frac{e}{h}\int_{0}^{\infty}d\epsilon|t(\epsilon)|^2
\Big[\Big(n_{F}(\epsilon -V)_{T+\Delta T}-n_{F}(\epsilon-V)_{T}\Big)-\Big( n_{F}(\epsilon+V)_{T+\Delta T}-n_{F}(\epsilon+V)_{T}\Big)\Big]\nonumber\\&&= N\frac{e}{h}K_{B}T\int_{0}^{\infty}dx|t(xK_{B}T)|^2
\Big[\Big[\Big(n_{F}(x -\frac{V}{K_{B}(T+\Delta T)}\Big)-\Big(n_{F}(x- \frac{V}{K_{B}T}\Big)\Big]\nonumber\\&&-\Big[\Big( n_{F}(x+ \frac{V}{K_{B}(T+\Delta T)}\Big)-n_{F}\Big(x+ \frac{V}{K_{B}T}\Big)\Big]\Big]\nonumber\\&&
\end{eqnarray}
This result  is presented in Figure 9. Comparing the thermoelectric current in Figure $7$  with Figure $9$, we observe that the reduction in the transmission function decreases the thermoelectric current. Figure  $9$ also show that at  small voltages bias the thermoelectric current is enhanced.

\vspace{0.2 in}

\textbf{VI- Proposed experimental set up for testing thermoelectricity }

\vspace{0.2in}

We will chose  a Weyl semimetal material with the surface at $z=0$  and  width few millimeters in the $z$ direction.To be in the mesoscopic regime the temperature should be  below $5k$  ,  the length in the $y$    direction $\approx 1 \mu m$ and the width in the $x$ direction below $ 0.1 \mu m$. On the left side of the sample we apply a heating device which will create a temperature $T+\Delta T $ The right side of the sample is held at temperature $T$.  A random magnetic field or paramagnetic impurities on the surface are requiered in order to observe  backscattering .The field is  $y$ dependent.  Backscattering will occur only when the right and left  channel have  the same  momentum  $k_{x}$.
As result,   we will observe a thermoelectric voltage between the left and right side of the sample. The sign of the voltage will depend on the   value of the  voltage $V$ bias. In the z direction we have a crystal of milimeter length.
 Our  computation predicts that for perfect transmission $ |t|^2=1$  (absence of backscattering) the thermoelectric cuurent is significantly  higher than the thermoelectric current obtained  when backscattering is allowed. This symmetry  afects the value of the thermoelectric current and can be used as a detection signature. At low  temperature we have only one channel , $ N=1$. With the increases  of the temperature the conductances will scale with the factor $N$, but the ratio between the  conductances   with and without backscattering  will be $N$  independent.

\vspace{0.2 in}

\textbf{VII-Conclusion}

\vspace{0.2 in}

Quantum effects are predicted   for thermoelectricity.  The computation is  performed for surface zero modes which are protected against backscattering. The effect of magnetic impurities allows to probe the thermal effect as function of the scattering length $l_{e}$ and   confirm the oscillation of the thermoelectric conductance as a function of the voltage bias. An  experimental set up  was proposed to test our theory.

\begin{figure}
\begin{center}
\includegraphics[width=4.5 in]{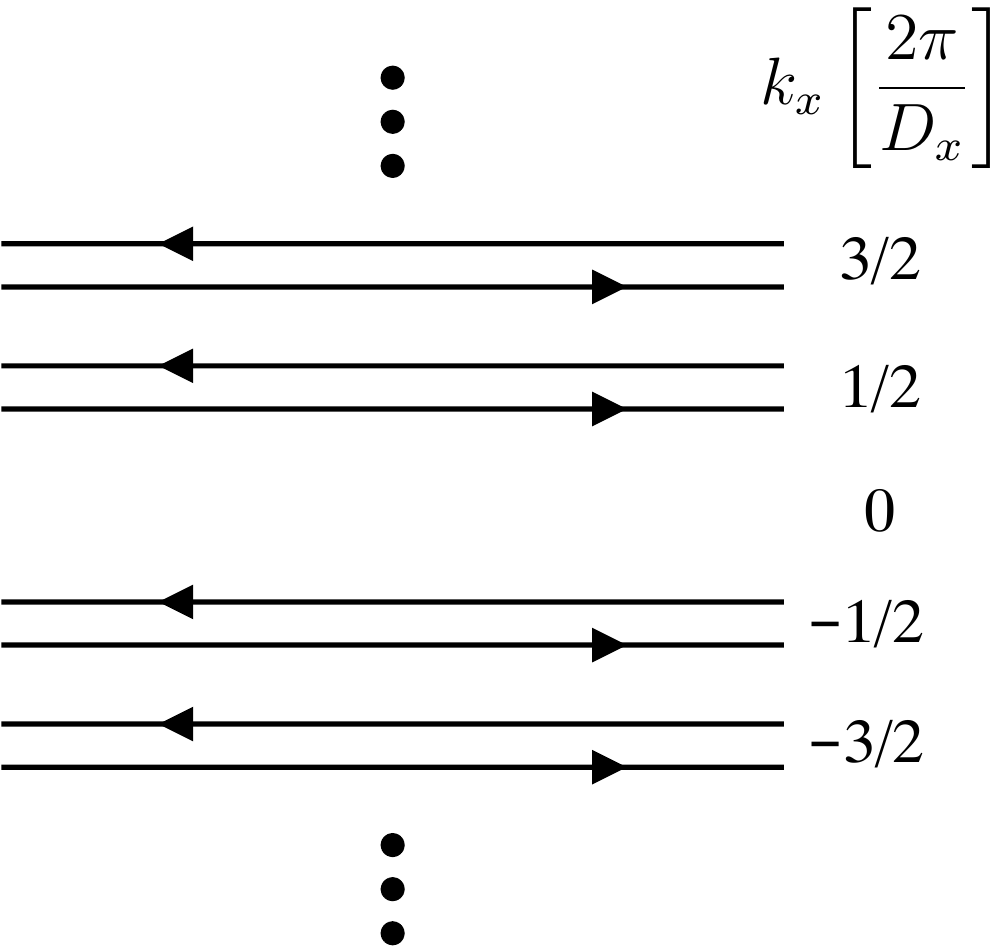}
\end{center}
\caption{The one dimensional propagating channels as a function of the momentum $k_{x}=\pm\frac{2\pi}{D_{x} }(n_{x}+\frac{1}{2})$. The figure shows a finite number of channels. We observe that for each momentum $k_{x}$ we have two conducting channels. }
\end{figure}

\begin{figure}
\begin{center}
\includegraphics[width=4.5 in]{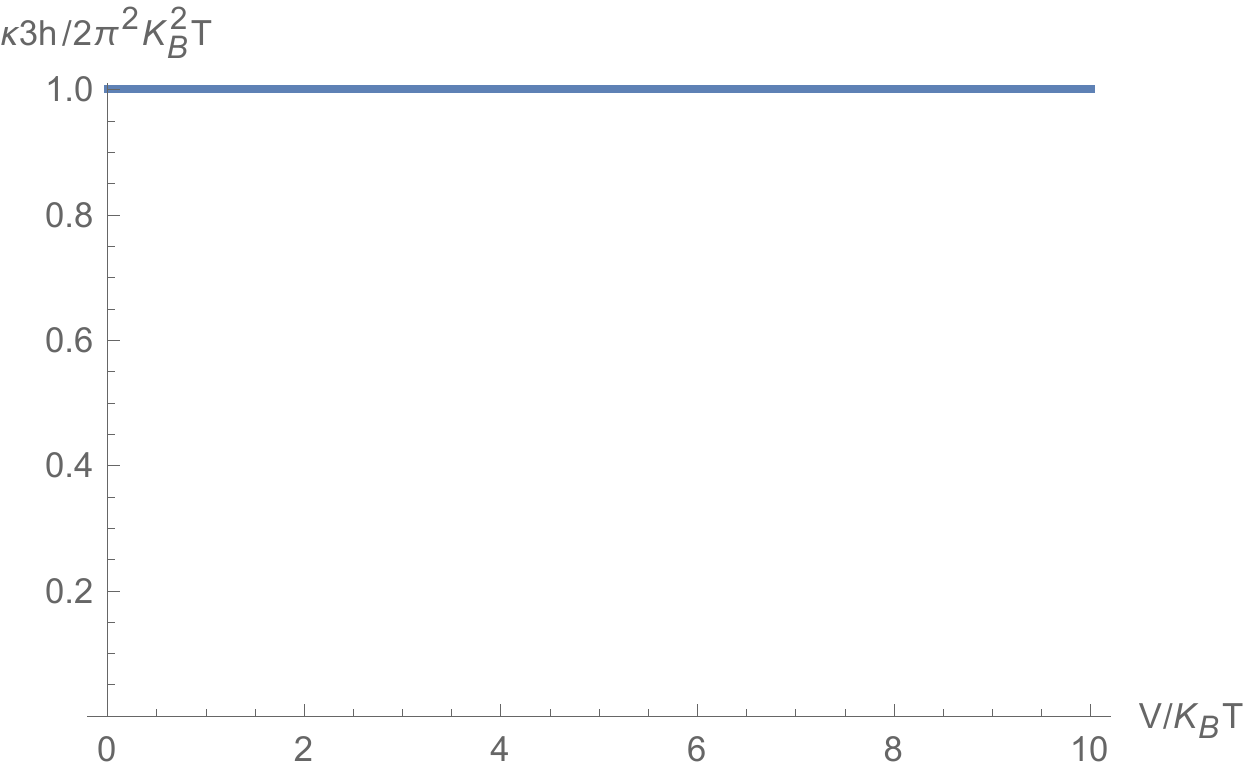}
\end{center}
\caption{The thermal   conductance    is  $\kappa=\frac{\pi^2K^{2}_{B}T}{3h}I[\frac{V}{K_{B}T}]$   where  $I[\frac{V}{K_{B}T}]\approx 1$, $|t|^2=1$}
\end{figure}

\begin{figure}
\begin{center}
\includegraphics[width=4.5 in]{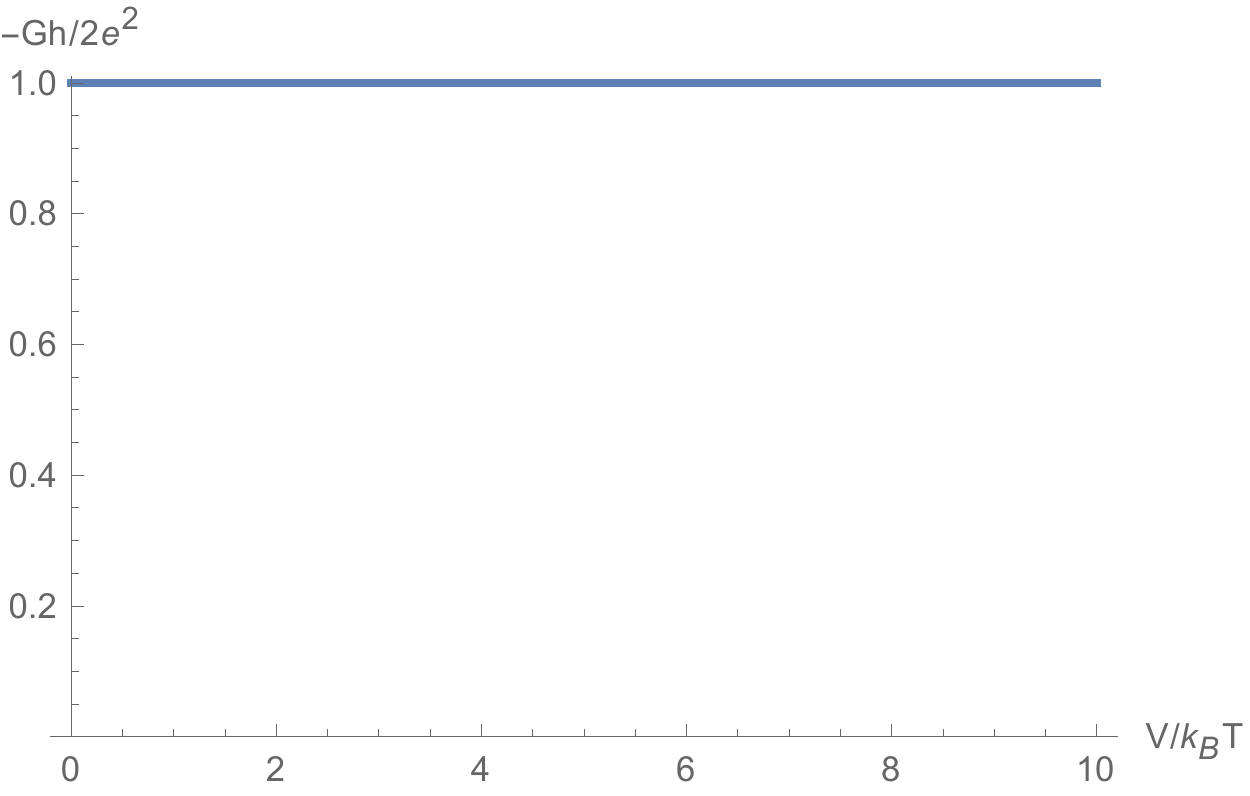}
\end{center}
\caption{The conductance         is given  by  $G=\frac{2e}{h}$.}
\end{figure}

\begin{figure}
\begin{center}
\includegraphics[width=4.5 in]{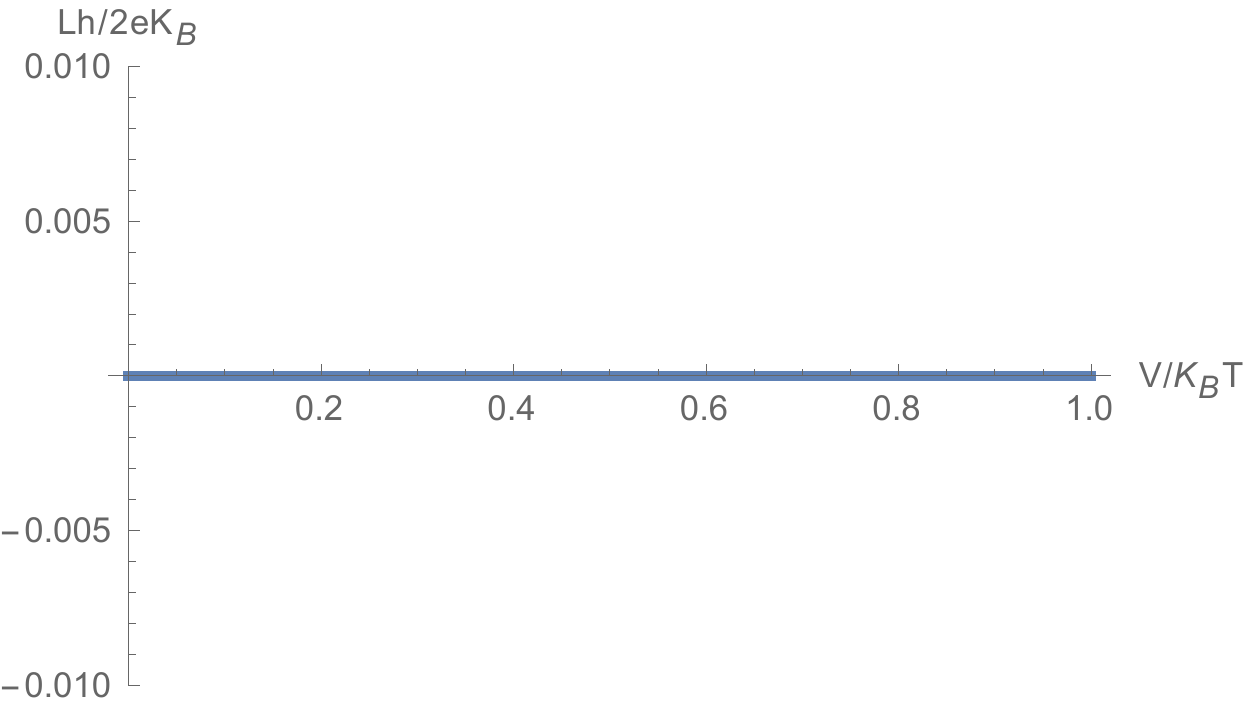}
\end{center}
\caption{The thermoelectric conductance $L=\frac{2e}{h}K_{B}$  for the transmission $|t(\epsilon)|^2=1$ }
\end{figure}

\begin{figure}
\begin{center}
\includegraphics[width=4.5 in]{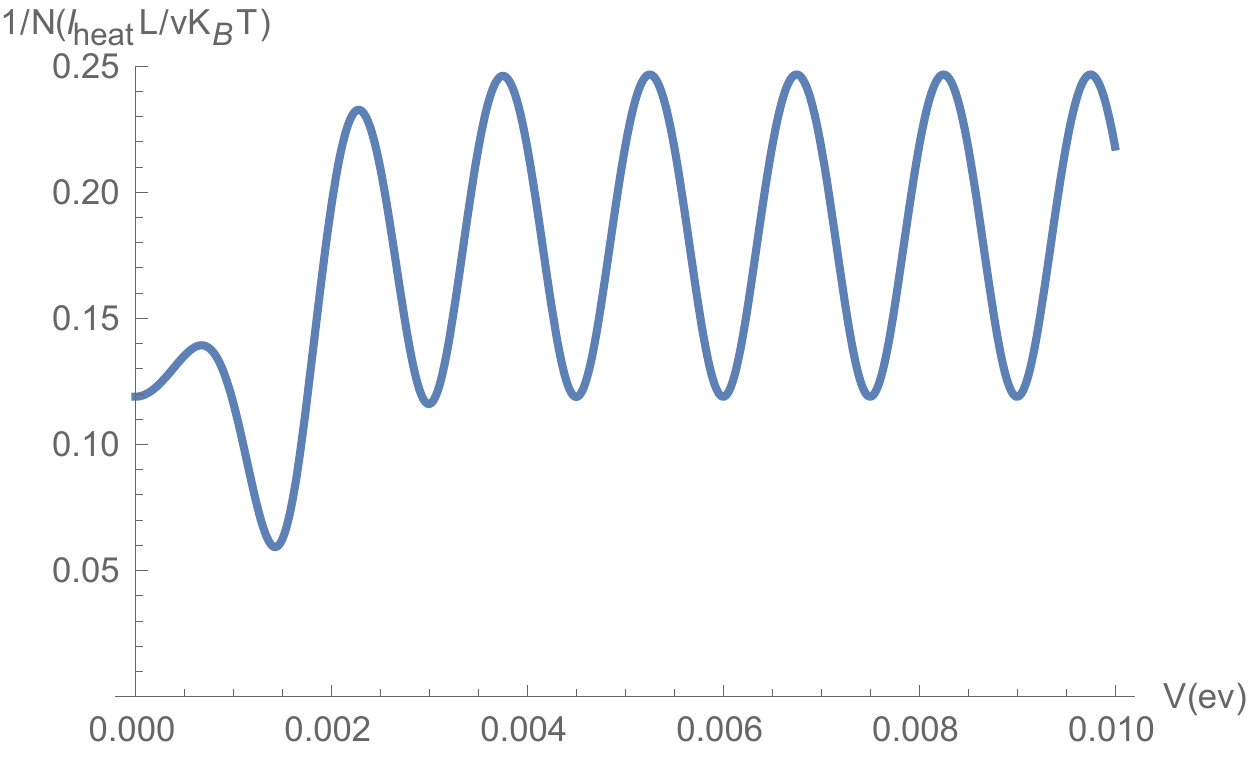}
\end{center}
\caption{ The thermal heat current plot is given for a temperature $T\approx 3$ (see figure $ 5$ )  as a function of the voltage bias $V= \hbar v(k_{0}- k_{F})$  ,the velocity was $10^5\frac{m}{sec}$ and  the temperature difference was  $0.1 K$ .This figure includes  ncludes particle and hole contributions . The heat current  fluctuates as a function of the bias voltage $V$ measured in electron volt, $I_{heat}=NK_{B} T \frac{v}{L}[0.15-0.25]\frac{ Joule}{sec}$}
\end{figure}

\begin{figure}
\begin{center}
\includegraphics[width=4.5 in]{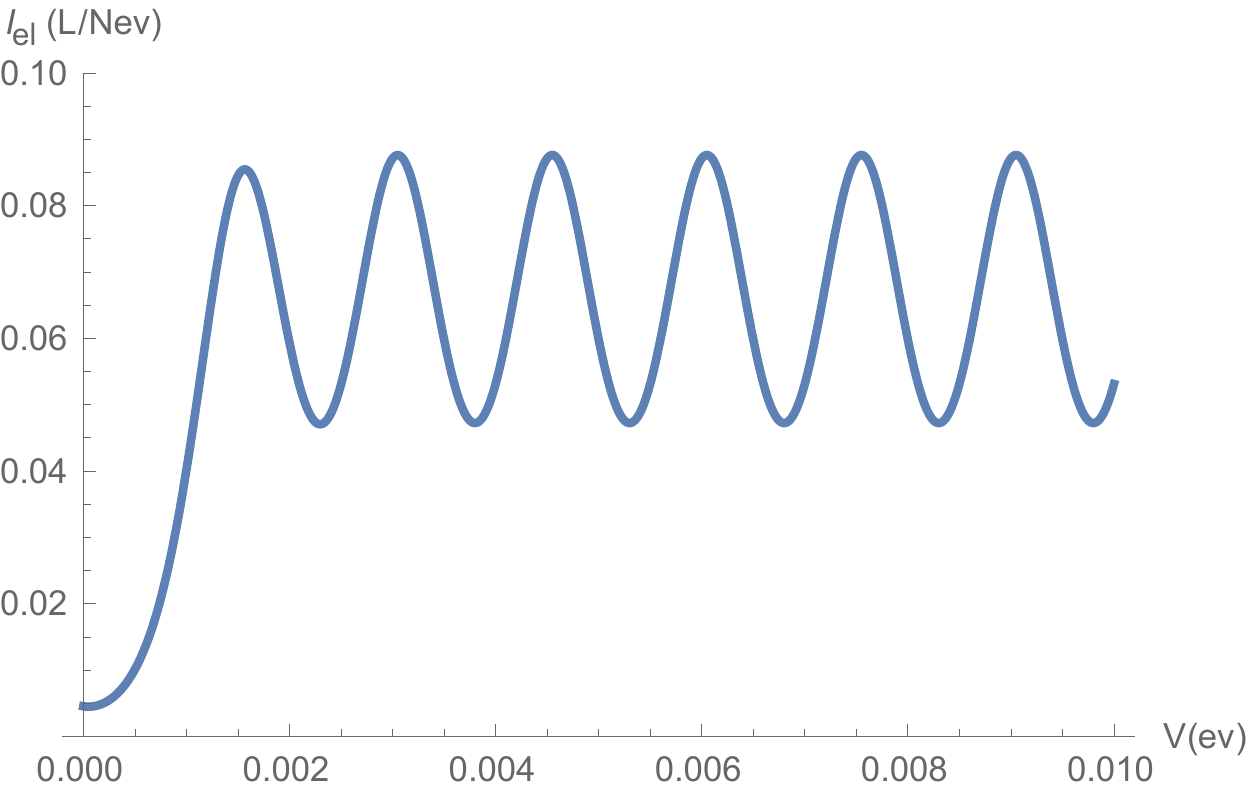}
\end{center}
\caption{The electrical conductance at $3k$ for $eV_{G}=0.0001ev$. $I_{el.}=N\cdot  \frac{ ev }{L}[0.04-0.08]Ampere.$}
\end{figure}

\begin{figure}
\begin{center}
\includegraphics[width=4.5 in]{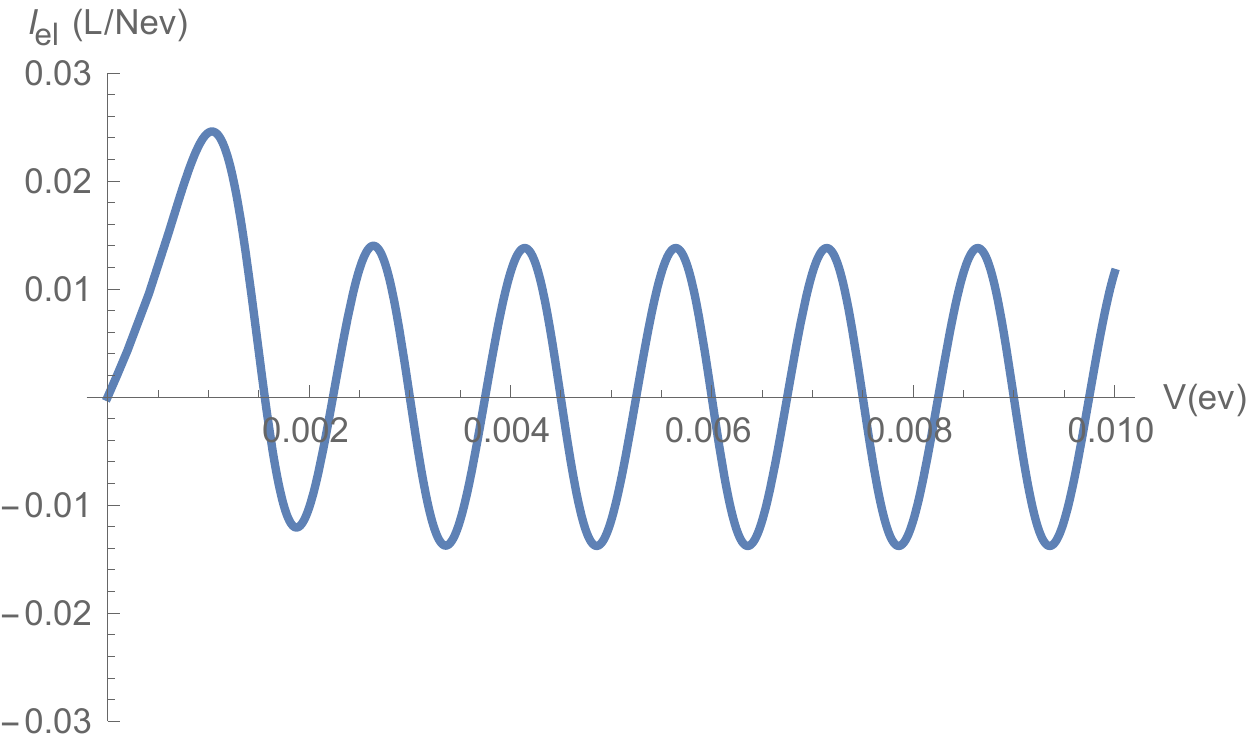}
\end{center}
\caption{ The Thermoelectric current for the transmission $|t(\epsilon)|^2=1$. $I_{el.}=N\cdot  \frac{ev}{L} ...\pm 0.02 Ampere$}
\end{figure}

\begin{figure} 
\begin{center}
\includegraphics[width=4.5 in]{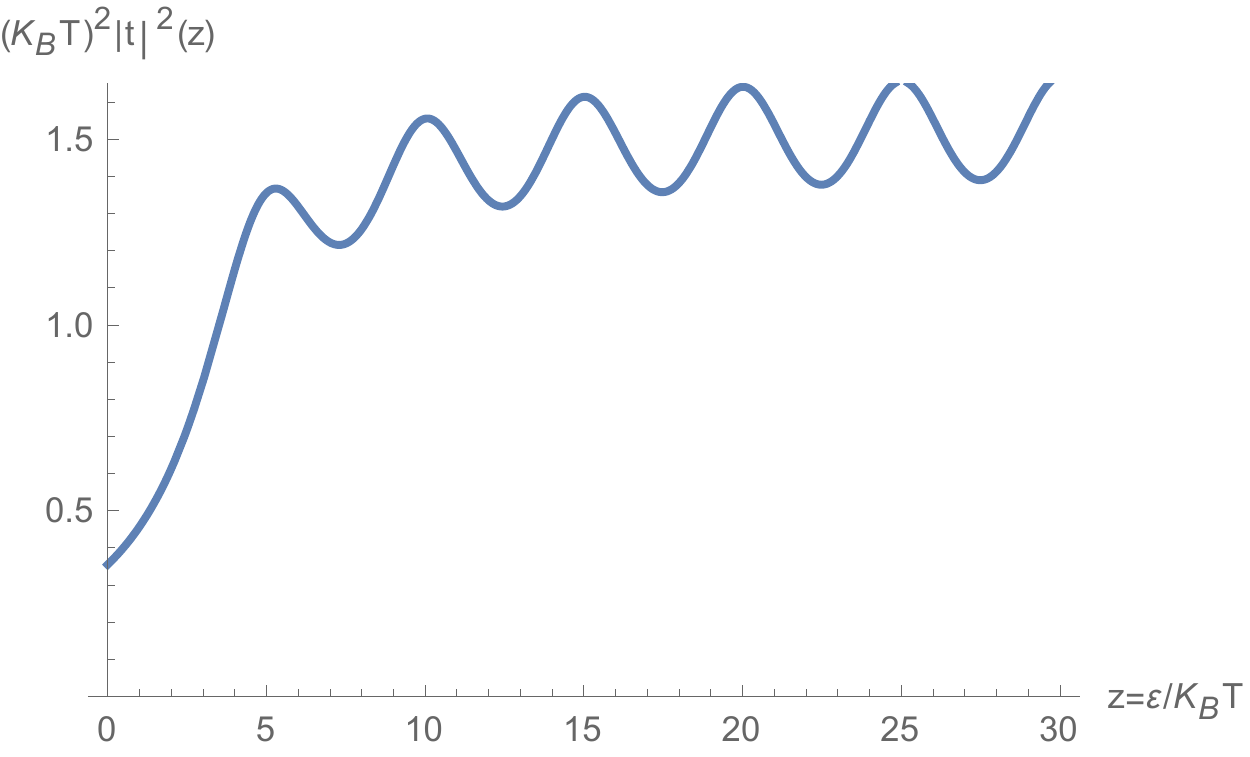}
\end{center}
\caption{ The transmission function   $|t(\epsilon)|^2\approx \sum _{n}\frac{(\frac{\Gamma}{2})^2}{(\epsilon -\epsilon_{n})^2+(\frac{\Gamma}{2})^2}$ .We show results   for  the case that $\epsilon_{0}\approx\Gamma$.}
\end{figure} 

\begin{figure}
\begin{center}
\includegraphics[width=4.5 in]{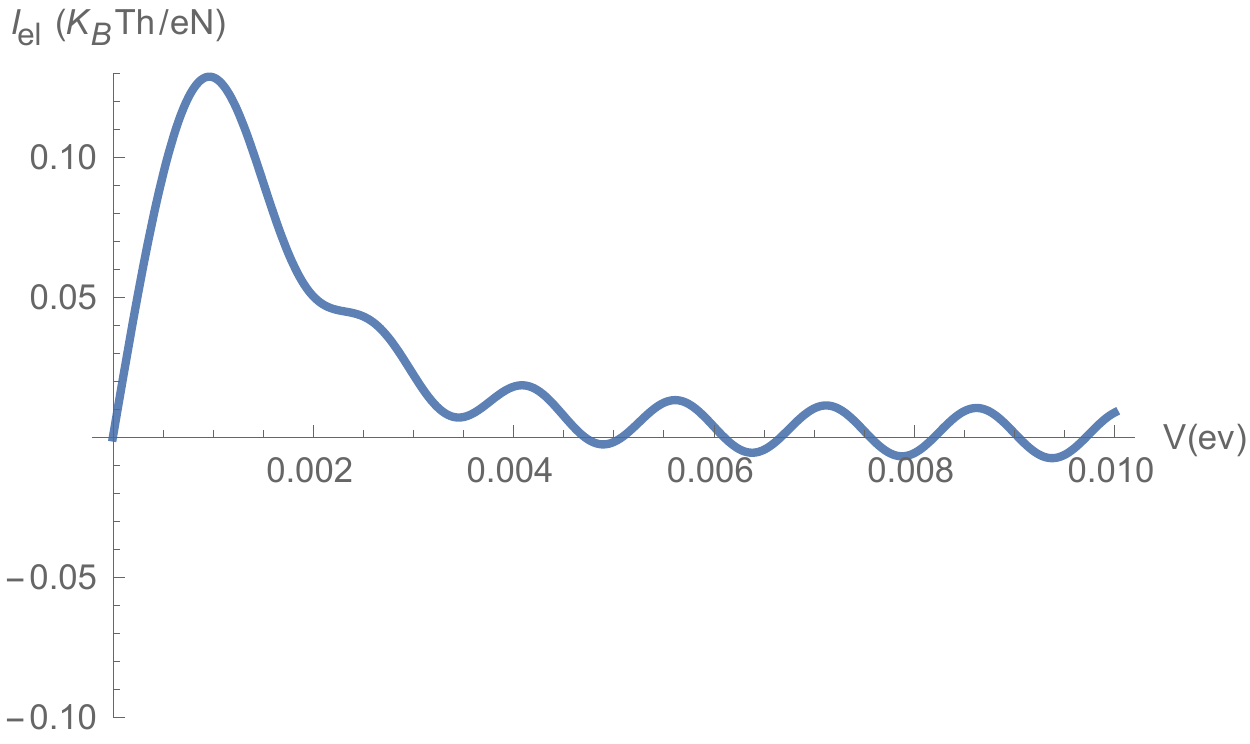}
\end{center}
\caption{The thermoelectric current  for the mesoscopic case  is   $I_{el.}=N\frac{eK_{B}T}{h}\pm 0.015....Ampere$ , $\delta T= 0.1 K$  at $T=3K$.} 

\end{figure}

\begin{figure}
\begin{center}
\includegraphics[width=4.5 in]{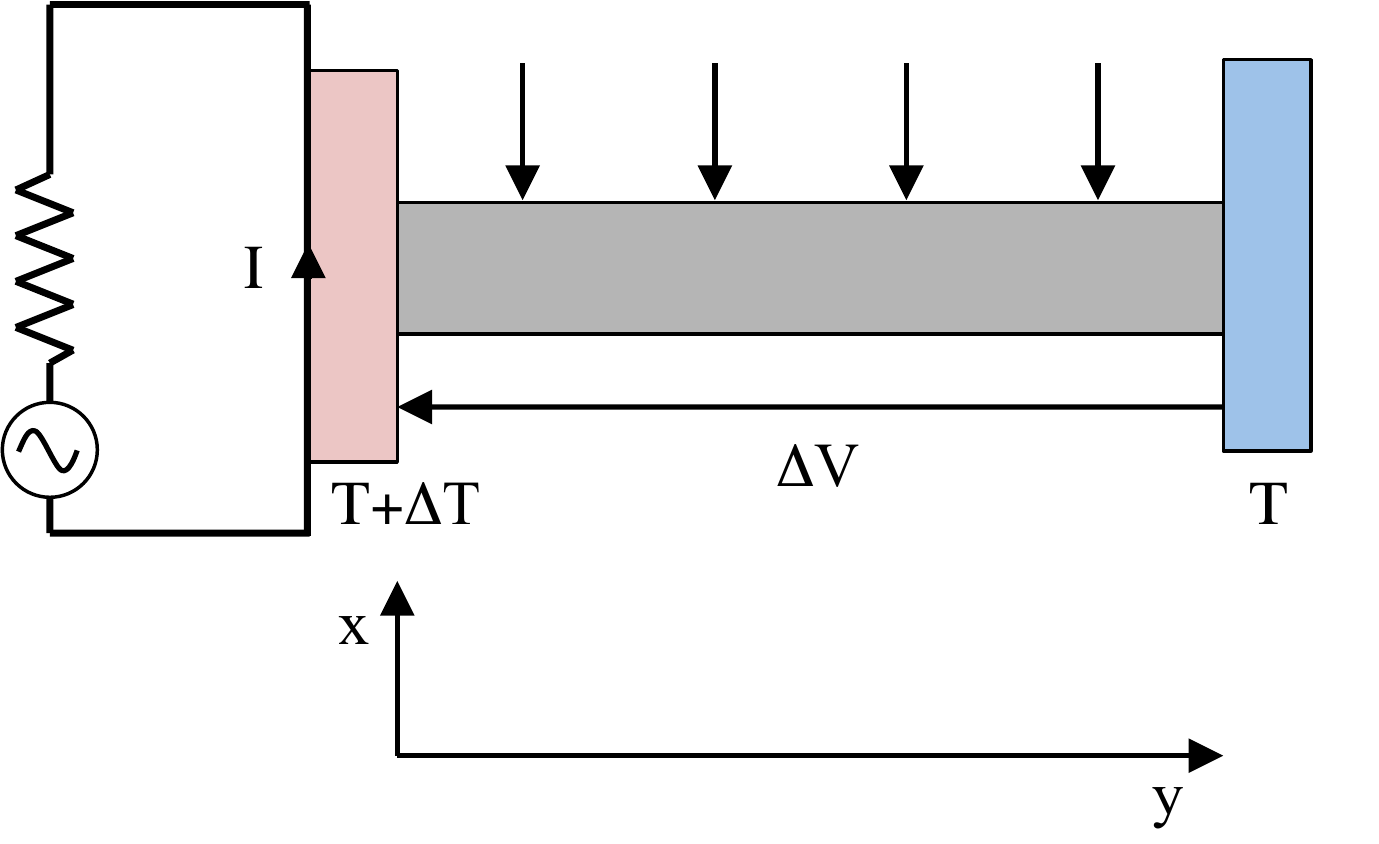}
\end{center}
\caption{The proposed experiment.The  effect of the paramagnetic impurities is showing by the arrow in the $x$ direction in the figure. We mention that the effect of the random paramagnetic impurities can be achived by applyng a magnetic field of $1 milimeter$  wavelength.The left side of the sample is connected to a current$ I$ to  create a elevated  temperature $T+\Delta T$ with trespect the right side. A voltage $ \Delta V$ is induced by the temperature difference }
\end{figure}
\end{document}